\documentclass{article}

\usepackage{natbib}
\usepackage[english]{babel} 
\usepackage{graphicx}
\usepackage{amsmath}
\usepackage{verbatim}
\usepackage{booktabs}
\usepackage{adjustbox}
\usepackage{rotating}
\usepackage{multicol}
\usepackage{hyperref}

\usepackage{color}
\definecolor{orange}{RGB}{255,127,0}

\usepackage{caption}
\usepackage{subcaption}

\usepackage{chngcntr}
\usepackage{pdflscape}
\usepackage{chngcntr}

\usepackage{geometry}
\usepackage[dvipsnames]{xcolor}

\title{Implied volatility smile dynamics in the presence of jumps}
\author{M. Magris, P. B\"{a}rholm, J. Kanniainen \\ 
{\small Laboratory of Industrial and Information Management, Tampere University of Technology, Tampere, Finland}}
\date{}

\pdfoutput=1

\begin{document}

\maketitle

\begin{abstract}
The main purpose of this work is to examine the behavior of the implied volatility smiles around jumps, contributing to the literature with a high-frequency analysis of the smile dynamics based on intra-day option data. From our high-frequency SPX S\&P500 index option dataset, we utilize the first three principal components to characterize the implied volatility smile and analyze its dynamics by the distribution of the scores' means and variances and other statistics for the first hour of the day, in scenarios where jumps are detected and not. Our analyses clearly suggest that changes in the volatility smiles have abnormal properties around jumps compared with the absence of jumps, regardless of maturity and type of the option. 
\end{abstract}

\section{Introduction}

In this paper, we examine how jumps in the underlying price affect the behaviour of the implied volatility smile. Our research is among the first analyses on the intra-day smile dynamics high-frequency option data. In fact, to the best of our knowledge, the intra-day dynamics of the implied volatility smile have \textit{not} been studied before, and certainly not in the context of jumps. This paper aims to fill this gap.

During the past years, the availability of high frequency data paved the way for new empirical research opportunities. The availability of intra-day option data allows us to study the dynamics of the volatility smile in a new perspective, in particular by taking the impact of sudden and large movements in the price of the underlying (jumps) into account. 

The analysis of the dynamics of the volatility smile is not a recent topic in financial econometrics and quantitative finance \citep[see e.g.][]{heynen1994} and several authors have suggested effective techniques to capture its movements including \citep{skiadopoulos2000,cont2002, fengler2003, benko2009,bernales2015}. The prevailing method to characterize the implied volatility smile is the principal component analysis (PCA) with possible variations. Existing methodological research constitutes a solid methodological background to our analyses based on intra-day option data. On the other hand, jumps in asset prices are an empirical fact \citep[e.g.][]{bates1996jumps, kanniain2016jump}) and often have been accommodated to extend stochastic volatility models. Several techniques are available to detect jumps in a timeseries, including  \citep{barndorff2004power, lee2008jumps}. In this paper, we exploit the well-received jump detection method introduced in \citep{lee2008jumps}. 

We provide statistical evidence that volatility smile dynamics is different in the presence of jumps in underlying returns in comparison to its \lq\lq normal\rq\rq dynamics without jumps. Most of the option pricing models, including \cite{bates1996jumps}, \cite{duffie2000transform} assume that jumps in stock price do not drive the stochastic diffusion of the variance process.\footnote{ The jumps of variance and price processes are often assumed to be correlated, but this should not be mixed with the commonly assumed independency between price jumps and continuous (brownian) component of variance process.} Our results are arguably controverting this independence assumption between jumps in the underlying price process and the volatility diffusion, suggesting a strong interconnection between them. 

The rest of the paper is organized as follows. 
In section \ref{SEC:THEORY} we review the concepts implied volatility smile and surface, as well as some of the techniques for the analysis of their dynamics. In section \ref{SEC:Data} we introduce our data, present the method used for jump detection and descriptive statistics for the principal components. Section \ref{SEC:PCBEHAV} studies the dynamics of the smile in terms of the distributions of mean and variance of the scores deducted from the principal components and in terms of their first moments. Finally section \ref{SEC:CONCL} concludes and suggests new possible research directions.

\section{Theorethical background}\label{SEC:THEORY}
\subsection{Implied volatility surface}
Since the introduction of the Black-Scholes theory as a standard reference framework for both academics and practitioners, the study and understanding of the implied volatility has been a major area of effort for financial econometrics \citep{gatheral2011volatility}. The Black-Scholes model assumes that implied volatility is constant, while empirically varies with respect to strike price $K$ and time to maturity $\tau$. Among the early researches \cite{rubinstein1994implied} found smile features in Black-Scholes IVs for the S\&P500 index options. \cite{xu1994magnitude} find the same features in the Philadelphia Exchange foreign currency option market and Heynen \citeyear{heynen1994empirical} in the European Options Exchange. 
\cite{dumas1998} points out that implied volatility decreases monotonically as thr strike price relative to the level of the underlying increases until $K/S = 1$, with the rate of decrease sharpening for shorter-maturity options. In a more recent study, \cite{lin2008smiling} investigate implied volatilities of individual equity and FTSE 100 index options and find the slope of the implied volatility smile to be significantly negative for both types of options. 
Moreover, they conclude that the smile for individual equity options is less pronounced than that of equity index options. Aside from these characteristics, the implied volatility also provides an alternative way of estimating the future volatility of an asset \citep{dumas1998, christensen1998relation, lamoureux1993forecasting, canina1993informational}. 

The smile and term structure features are indeed merely cross-sections of the so-called implied volatility \textit{surface} (IVS) that jointly describes the relationship of the implied volatility (IV) with different strikes and maturities for a given time (surface in the left plot of Figure \ref{fig:right}).
\cite{cont2002} define the IVS as the function
\begin{displaymath}
\sigma_{t}^{BS}:(K,\tau) \rightarrow \sigma_{t}^{BS}(K,\tau) \nolinebreak
\end{displaymath}
mapping a point $(K,\tau)$ to a point on the surface $ \sigma_{t}^{BS}(K,\tau) $. 
This expression, reveals the three dimensions in which implied volatility varies: strike price $K$, time-to-maturity $\tau$ and time $ t $. Variations in the dimensions $K$ and $\tau$ are referred to as implied volatility statics, whereas variations in the dimension $t$ are referred to as implied volatility dynamics.

Considering briefly the dynamics of the implied volatility surface, option traders often employ simplistic rules of thumb for depicting the evolution of the IVS through time. In \cite{cont2002} the authors focus on the so-called \lq\lq sticky moneyness\rq\rq and \lq\lq sticky strike\rq\rq rules for estimating the evolution of the IVS through time, pointing out that these rules of thumb are merely deterministic laws of motion for the IVS and not realistic. Essentially two approaches exist in literature addressing the issue of the non-flat profile and time variation of the IVS. The first approach explains the IVS and its profile by allowing additional degrees of freedom into the Black-Scholes model. These degrees of freedom are introduced into the model by allowing stochastic volatility, allowing jumps in the underlying price or combinations of these two \citep[see e.g][]{gatheral2011volatility}.  \cite{skiadopoulos2001} and \cite{cont2002} underline several problems in seeking this approach. These complications suggest for a completely reversal approach where the starting point is the IVS, which is taken as given and used for attaining information on the underlying process \citep{skiadopoulos2001}.

\subsection{Characterization of the implied volatility surface}\label{SUBSEC:CHARIVS}
\cite{skiadopoulos2000} explore how many factors are needed to model the dynamics of the implied volatility surface (IVS) and how these can be interpreted. The technique they use to shed light on these questions is the principal components analysis (PCA), which can be considered the method of choice in literature to answer questions related to the dynamical aspects of the IVS. It is common practice to identify the number and sources of shocks that move, for example, at-the-money implied volatilities by principal components analysis \citep{fengler2003}. This is exactly the approach employed by \cite{skiadopoulos2000}, who use daily data on futures options on the S\&P500 index and study the dynamics of the IVS by forming maturity buckets across which the volatility smiles are averaged and then applied PCA to. They extract two principal components that interpret as a parallel shift of the surface and Z-shaped twist of the surface. The extracted components explain, on average, 60\% of the variation of the surface. \cite{panigirtzoglou2004} also extract only two principal components in their study of the dynamics of option price implied probability distributions. Some authors \citep[e.g][]{cont2002, fengler2003} however conclude that three factors are needed to capture an adequate amount of the total variation of the IVS to satisfactorily model its evolution.
\cite{fengler2003} also studies the dynamics of IVs but remarks that the commonly applied method also employed by \citep{skiadopoulos2000} is lacking because it neglects the surface structure of the implied volatilities.

The relationship between standard PCA and the so-called common principal component analysis (CPC) is that PCA is a dimension reduction method for one group of data \citep{fengler2006}, whereas CPC is a dimension reduction method for multiple groups of data. This relationship is essential when these methods are applied to study the IVS, because the IVS consists of implied volatilities for different maturity groups.
\cite{fengler2003} thus perform their study using CPC where the space spanned by the principal components is identical across maturity slices. \cite{fengler2006} shows that the eigenvectors, or principal components, in fact are close to identical across different maturities of the IVS, justifying the use of this more general version of PCA in studying the IVS.

Recognizing the problems associated with the studies described above, \cite{cont2002} apply a slightly modified approach in order to actually analyse the joint dynamics of the entire IVS. Applying this analysis to S\&P 500 and FTSE 100 index options data, the authors extract three principal components that can satisfactorily explain the variance in the IVS. They also propose a three-factor model of the form 
\begin{equation*}
\textrm{log}I_t(m,\tau) = X_t(m,\tau) = X_0(m,\tau) + \sum_{k=1}^{3}x_k(t)f_k(m,\tau) \nolinebreak ,
\end{equation*} 
where $ m= K/S_t$ is the moneyness, $ X_0(m,\tau) $ is the surface at present time, $ f_k(m,\tau) $ are the principal components and $x_k$ are stochastic processes driven by independent sources of noise. Note that the sticky moneyness rule described previously is a special case of this model, where $x_k$ are zero.

The analysis of the dynamics of the IVS has sequentially relied on a number of different methods. First applying PCA to at-the-money ($m=1$) implied volatility term structures \citep[see e.g][]{heynen1994,zhu1997} and single smiles for a fixed maturity \citep{skiadopoulos2000}.
Second, applying common principal components analysis to maturity slices of the IVS \citep[see e.g][]{fengler2003} to account for the surface structure previously neglected.
Finally, in the more recent research the point of view is shifted to the functional case, where the observed IVS is taken as a single object  (namely a function) instead of a sequence of separate observations at chosen time-to-maturities and moneyness \citep{fengler2006}. This shift allows the use of techniques that enable the analysis of the whole surface. \cite{cont2002} on their work based, upon the others on \citep{balland2002}, applied functional principal components analysis (FPCA) in the context of the IVS.

Since the wide literature exploiting the standard PCA and because of the nature of our analyses, which addresses the dynamics of the smile in presence of jumps in the underlying, in this research we are conveniently adopting the well-known standard PCA analysis, for which a short description is provided below. Suppose that $ \mathbf{X} \in \mathbf{R}^{n, p}$ is a data matrix of $p$ random variables for $n$ observations, $\mathbf{X} = (\mathbf{x}_1, \ldots , \mathbf{x}_p) $. In our case, $\mathbf{x}_i$ are the observations of the implied volatilities in the $i$-th moneyness bin. PCA replaces the set of $p$ correlated and unordered variables by a set of $ k \leq p $ uncorrelated and ordered linear projections $\mathbf{z}_1, \ldots ,\mathbf{z}_k$ of the original variables \citep{izenman2008}. The linear projections can be written as:
\begin{equation}\label{EQ:LINPROJ}
\mathbf{z}_j = \mathbf{b}{_j}\mathbf{X}^T = b_{j1}\mathbf{x}_1 + \ldots + b_{jp}\mathbf{x}_p, \qquad j = 1,2, \ldots ,k \nolinebreak ,
\end{equation}
where $ \textbf{b}_j $ is the vector of loadings for the $j$-th component. The goal is to find the projections that minimize the loss of information. 
When the coefficient vectors $ \mathbf{b}_j$ are picked so that the projections $z_j$ are ranked in decreasing order of variance, and that $z_j$ is uncorrelated with all the $z_i$ (for $ i < j $), we call the linear projections of equation \eqref{EQ:LINPROJ} as the $j$-th principal components of $\mathbf{X}$. 

Given the $p \times k$ loading matrix $\textbf{B} = \left( \mathbf{b}_1^T, \ldots, \mathbf{b}_k^T \right)$ of the first $k$ principal components, the elements of the $n	\times k $ matrix $\mathbf{S} = \mathbf{XB}$ representing the data matrix $\mathbf{X}$ on the principal components space, are commonly called \textit{scores}. The $j$-th column of $\mathbf{S}$ collects the scores associated with the $j$-th principal component. For further details and proofs see e.g. \citep{izenman2008}.

\section{Data}\label{SEC:Data}
\subsection{Intra-day option data}
\begin{figure}[!ht]
\centering
\includegraphics[scale=0.5]{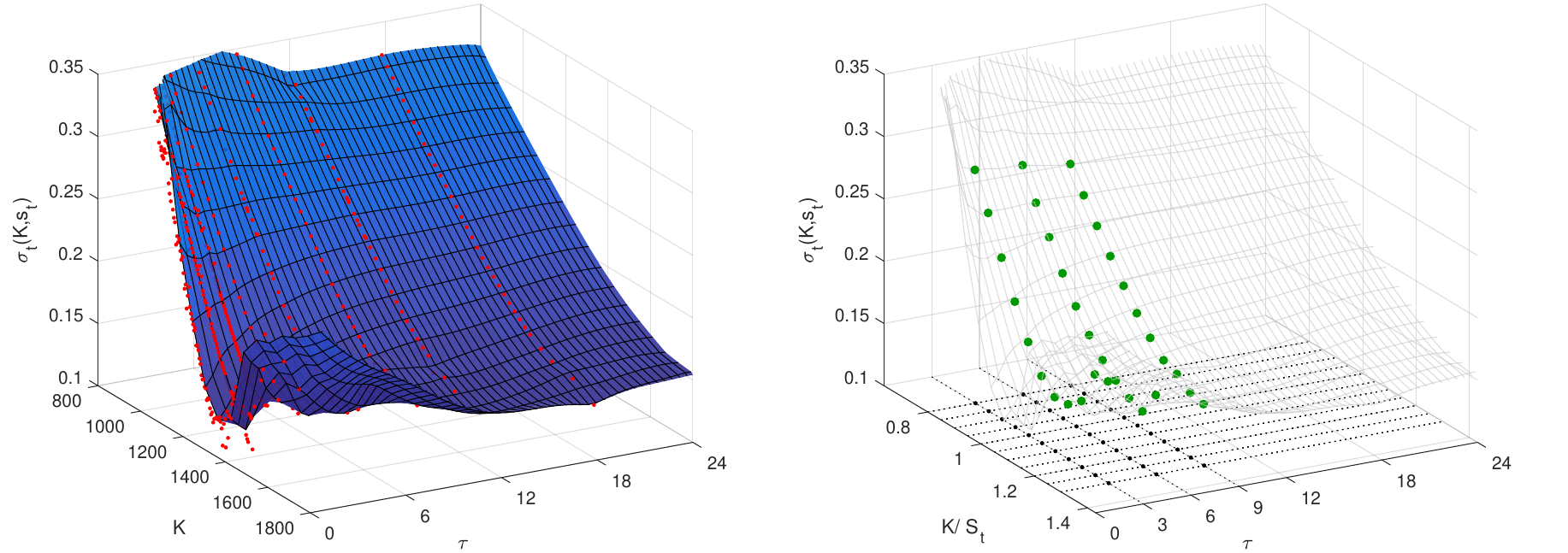}
\caption{IVS construction and smiles extraction (30-Dec-2010 at 15:20:00, the underlying price is 1258.57\$). In the left panel implied volatilities obtained from market prices (red dots) are used to fit the IVS. The right panel illustrates the three selected maturities ($\tau$ expressed in months), the adopted moneyness range and bins (black dots on the moneyness-maturity plane) used in our analyses to slice the surface and extract the sampled smiles for PCA (green dots).}
\label{fig:right}
\end{figure}

We study the dynamics of the volatility smile implied from the market prices of call and put out-of-money and at-the-money options on the S\&P500 index, SPX. The dataset studied contains the spot price and the cross-sections of put and call prices for SPX on one minute interval and spans a five-year period from the beginning of 2006 to the end of 2010 (1,259 trading days in total)\footnote{The option data is provided by CBOE Livevol.}. Using this data, Black-Scholes implied volatility is computed for each one of the available points on the time to maturity-strike plane for every minute, on every trading day between 09:31:00 and 16:15:00 in the time 2006-2010. The exclusion from our analyses of the in-the-money options and the use of out-of-the-money (OTM) and at-the-money (ATM) puts and calls only, is motivated by the fact that OTM and ATM options are those of most interest since traded the most and thus liquid. Also, VIX is calculated excluding in-the-money options. We further obtain the implied volatility surface, in this study via thin plate spline interpolation \citep{wahba1990spline} and select specific slices of time to maturity for fixed moneyness ranges.\footnote{\cite{homescu2011} lists a number of practical reasons for obtaining a smooth surface: the discrete market prices at different strikes and maturities pose the problem of how to construct a continuous IVS.} Figure \ref{fig:right} illustrates this procedure for a specific epoch. The left plot illustrates the surface construction by interpolation applied to the implied volatilities computed from market prices, while the right plot illustrates the selection of the moneyness range and bins for three specific times to maturity (3, 6 and 9 months) and the corresponding smiles extracted from the surface. 

A necessary transformation to impose on the data in order to facilitate the analyses, is to shift from the absolute coordinates of strike price $ K $ to relative coordinates of moneyness $ m = K/S_t $, where $ S_t $ denotes the spot price of the underlying at time $ t $. 
This stems from the variability of $ K $ as $ S_t $ fluctuates: in absolute coordinates, we would not be able to construct strike variables that would allow principal component analysis due to the shifting of $ K $ with the shifting of $ S_t $. The moneyness range where the dynamics is analyzed is restricted to the interval $ \left[ 0.8, 1.3 \right] $.
This choice is a trade-off between two competing goals. First, we would like to be able to study a wide range of moneyness in order to make conclusions about deep-in-the-money ($ m \gg 1$ for OTM calls) as well as deep-out-of-the-money ($ m \ll 1$ for OTM puts) implied volatility dynamics. However, the width of the moneyness range is restricted by numerical uncertainties related to the surface smoothing and low liquidity at extreme moneyness values \citep{cont2002}. As seen in the left plot in Figure \ref{fig:right}, the number of implied volatilities observed at a given time is rather limited and heavily concentrated on the lower end of the time to maturity spectrum.
The range utilized in this study can be considered sufficiently wide for meaningful analyses since it yields 10 variables with a moneyness bin width of 0.05 and includes the at-the-money point $m = 1$, which is generally the most liquid moneyness for exchange traded options.

\subsection{Detecting jumps in the spot price data of SPX}\label{SUBSEC:Jumps}

The method of jump detection implemented in this research is that of \citep{lee2008jumps}: a nonparametric test applicable to a wide number of financial time series, provided that high-frequency data is available. In the present study, this method is used to detect jumps in the price of the underlying (which is available for every minute). In \citep{lee2008jumps} the test statistic $\mathcal{L}$ is based on returns that are scaled by realized bipower variation. The test statistics is shown to be approximately normally distributed when the underlying log-price comes from a standard brownian motion ($H_0$), which does not allow for jumps. For a given significance level, a threshold $\beta^*$ is derived such that when the observed test statistics is greater $\beta^*$ the null hypothesis is rejected (it's unlikely that the observed log-price path comes from a pure jump-free brownian motion, $H_1$).   For more information about the jump detection method, we refer \citep{lee2008jumps}.

Jump detection applied to our SPX option data with a detection window of 5 minutes leads to 396 jumps in total.
Due to the concentration of jumps in the first hour of the trading day ($\sim 85\%$), this paper opts to ignore jumps that are detected after 10:31:00, and study the dynamics of smiles only in the mornings where jumps are detected (between 9:31:00 and 10:30:00) in comparison with mornings of days when no jumps are detected.
In this way, we obtain a sample of 338 jumps. Due to multiple jumps in the same morning the total amount of days showing one or more jumps in the first hour is 290. At the same time, we constructed a subset of days where no jumps have been detected, against which the smile behaviour on mornings with detected jumps is compared.
Thus in the comparison data sample we only consider the remaining 940 days with no jump detected in the first hour.
For the sake of robustness, we implement the test statistic of \citep{lee2008jumps} using different sampling intervals for the price of the underlying, in particular we also adopted 15 minutes interval: 80 jumps detected in total, 57 of them in the first hour, for a total number of 54 days with jumps. The number of non-jump days is still 940 (indeed, we take as no-jump day any day showing no jumps in neither 5 nor 15 minutes detection windows).

\subsection{Implied volatility smile characterization by the first three principal components}\label{SC:Characterization}

As earlier reviewed, previous studies have extracted two or three principal components explaining typically around 80\% to 90\% of total variance in the implied volatility smile. In this study, we extract three principal components and study the dynamics of the first difference of the implied volatility ($\Delta IV$). In fact, it has been a common practice in the field not to deal with the IV itself but rather with its changes \cite[e.g.][]{skiadopoulos2000,fengler2003,panigirtzoglou2004, cont2002}. Therefore we analyse the \textit{changes} in implied volatility between consecutive times ($\Delta IV_t = IV_t - IV_{t-1}$), where $IV$ refers to implied volatility.

Table \ref{TAB:PERCVAR} underlines that the loss of information (unexplained variance) when characterizing the smile with these components is quite moderate \citep[supporting the methodologies adopted in][]{cont2002,fengler2003}.


\begin{table}[!htb]
\centering
\scalebox{1}{
\begin{tabular}{lcccc}
\toprule
{Maturity} & PC1 & PC2 & PC3 & Total \\  \midrule
{3 months}  &   65.3\% & 10.3\% & 9.4\%  &  85.0\% \\
{6 months } &   59.8\% & 15.2\% & 11.8\% &  86.9\% \\
{9 months}  &   54.7\% & 19.2\% & 12.7\% &  86.7\% \\
\bottomrule
\end{tabular}}
\caption{Percentage of variance explained by the first three principal components.}
\label{TAB:PERCVAR}
\end{table}

\noindent Note that slicing the surface in the direction of the moneyness dimension corresponds to extracting IV smiles for fixed maturities. Our choice of dealing with 3, 6 and 9 months maturities is motivated by the fact that wider maturity ranges correspond to lower liquidity scenarios. 

As earlier introduced, numerous authors have applied PCA-related techniques to study the dynamics of the implied volatility smile and the interpretations for the components they have extracted are remarkably consistent. However, to the best of knowledge of the authors, no studies have been conducted where PCA would have been applied to study the \textit{intra-day} dynamics of the smile. Indeed, previous research has only utilized data sampled at daily intervals. Therefore, it is considered worthwhile to study whether the interpretations of the components remain roughly the same when considering intra-day data. Furthermore, the interpretability of the PC we obtain is crucial for making meaningful conclusions about the smile behavior around jumps with respect to their behavior when there are no jumps.

\begin{figure}[!ht]
\centering
\includegraphics[scale=0.75]{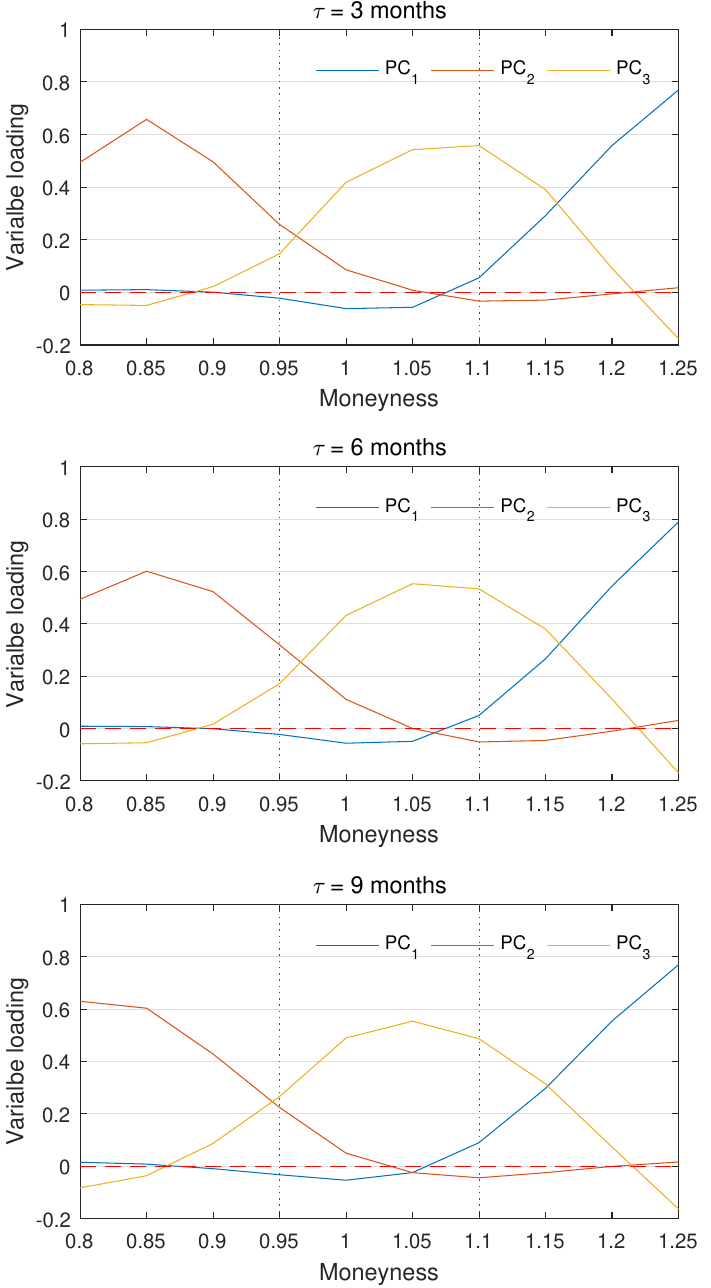}
\caption{Varimax-rotated variable loadings of the first three principal components for differences in implied volatility ($\Delta IV$). The red vertical  dotted lines approximately define the OTM puts region (leftmost area), ATM options region (central area), OTM calls region (rightmost area).}
\label{fig:loadings}
\end{figure}

The principal components can be interpreted by inspecting what are commonly called parallel coordinate plots, where the loadings of each principal component are plotted against the indices of the original variables (Figure \ref{fig:loadings}). Loadings tell how much each original variable contributes to a principal component. Commonly (Varimax) rotation is applied to the original principal components in order to facilitate their interpretation, while preserving their uncorrelation. This method generally yields to components that have a clear interpretation in terms of the original variables.

Figure \ref{fig:loadings} can be inspected to seek an interpretation for the three components.
For the 3-months maturity, the first PC (blue line) is interpreted as representing out-of-money (OTM) call options, because the bins of moneyness greater than one are highly loaded on this component, while the other two components exhibit lower loadings in the same range.
The second PC (red line) is highly loaded in the moneyness bins corresponding to out-of-the-money put option. Similarly, the third PC (yellow line) is interpreatable as at-the-money (put and call) options (ATM). Moving to the other maturities, we notice that there are no discrepancies with the previous case: the interpretations for the components are unchanged.

Based on Figure \ref{fig:loadings}, the principal components are renamed according to their interpretations. Henceforth, the principal components will be often referred to by the following interpretations: PC1 -- OTC Call, PC2 -- OTM Put, and PC3 -- ATM.

The principal component data was inspected for intra-day seasonality: we found that, on average, the implied volatility is considerably higher during the first minutes of the trading day. To remove this effect, the means of all the three retained principal components for each observation minute across all trading days in the data sample were subtracted from each observation of the corresponding minute.\\

\section{Behaviour of the implied volatility smile in the presence of jumps in the underlying}\label{SEC:PCBEHAV}

In this section, we study the dynamics of the implied volatility smile around jumps. 
In the following one, we report and discuss the analyses for the changes of the implied volatilities $\Delta IV$ and considering a jump detection window of 5 min. However, for robustness purposes, results for 15 min windows have been implemented as well. 

Being this the first research to tackle the problem of the smile dynamics around jumps, there is no existing methodological background we can consider for our research purpose. Therefore, we set up the methodology described in the next subsection aimed at detecting whether the scores implied by PCA may be on average different where jumps occur and when not, and so $\Delta IV$. This difference is in this work assessed (i) in terms of changes in the distribution of the scores means and variances in the first hour of the day and (ii) in terms of changes in their respective means.

\subsection{Methodology and notation}
For a given principal component, we first compute the scores for the first 60 minutes of each day differentiating for days with jumps and days without jumps in the first trading hour. In the following, index will $j$ denote variables and quantities related and dependent (directly or not) to the subsample of days containing jumps in the first hour ($j = 1$), and not ($j = 0$). 

We thus collect the scores in two distinct matrices $S_j$ where the generic row $s^j_d$ contains the scores of the first hour of day $d$, $s^j_d  =  \left\lbrace s^j_{d,1}, ..., s^j_{d,60} \right\rbrace$.
For each day, from $s^j_d$ we extract (i) the mean score of the first hour $\mu_{dj}$ and (ii) the variance of the scores in the first hour $\nu_{dj}$. We denote the respective samples with $M_j$ and $\Sigma_j$ (observations below the 2\% and above the 98\% quantiles have been removed to prevent from biases related to outliers).

\begin{table}[!ht]
\centering
\begin{tabular}{ l c c c c c c c }

\cline{2-5} \cline{7-8}
  & \multicolumn{1}{|c}{minute\textsubscript{1}} & ... & ... & \multicolumn{1}{c|}{minute\textsubscript{60}} & & \multicolumn{1}{|c|}{$M_j$} & \multicolumn{1}{c|}{$\Sigma_j$}\\
\cline{1-5} \cline{7-8}
\multicolumn{1}{|c|}{Day 1} 	 & $s^j_{1,1}$ & ... & ...	& \multicolumn{1}{c|}{$s^j_{1,60}$} 	& $\rightarrow$ &\multicolumn{1}{|c|}{$\mu_{1j}$} & \multicolumn{1}{c|}{$\nu_{1j}$}\\
\multicolumn{1}{|c|}{...}   	 & ... & ... & ...			& \multicolumn{1}{c|}{...}				& $\rightarrow$ &\multicolumn{1}{|c|}{...} 		&\multicolumn{1}{c|}{...}\\
\multicolumn{1}{|c|}{Day d} 	 & $s^j_{d,1}$ & ... & ...	& \multicolumn{1}{c|}{$s^j_{d,60}$} 	& $\rightarrow$ &\multicolumn{1}{|c|}{$\mu_{dj}$} & \multicolumn{1}{c|}{$\nu_{dj}$}\\
\multicolumn{1}{|c|}{...}  	 	 & ... & ... & ...			& \multicolumn{1}{c|}{...}				& $\rightarrow$ &\multicolumn{1}{|c|}{...} 		&\multicolumn{1}{c|}{...}\\
\multicolumn{1}{|c|}{Day $n_j$}  & $s^j_{n_j,1}$ & ... & ...& \multicolumn{1}{c|}{$s^j_{n_j,60}$} 	& $\rightarrow$ &\multicolumn{1}{|c|}{$\mu_{n_jj}$} &\multicolumn{1}{c|}{ $\nu_{n_jj}$}\\
\cline{1-5} \cline{7-8}
\end{tabular}
\caption{Representation of the work flow. For each day ($d$) in the scores matrix $S_j$ we consider the mean ($\mu_{dj}$) and the variance ($\nu_{dj}$) of the scores in the first hour. $M_j$ and $\Sigma_j$ represent the respective data vectors whose distributions we analyze.}
\end{table}

Based on the empirical distributions of $M_j$ and $\Sigma_j$ we investigate the change in implied volatility around jumps with the following two methods:
\begin{itemize}
\item[i)] The two-sample Kolmogorov-Smirnov test is used to deduce whether means and variances of the principal components in mornings with jumps in the underlying are compatible with their respective distributions on normal days without jumps. This corresponds to test for statistically significant differences in the distributions of $M_0$ and $M_1$ and differences in the distributions of $\Sigma_0$ and $\Sigma_1$ (distributions around jumps versus data around non-jumps).
\item[ii)] The Welch-U test is used to compare the means of $M_0$ and $M_1$ and means $\Sigma_0$ of and $\Sigma_1$. This corresponds to test whether there's statistical evidence of differences in the mean values of the ranks.
\end{itemize}

\noindent All the above presentation applies and is implemented for each of the three principal components.

\subsection{Two-Sample Kolmogorov-Smirnov Test}\label{SC:KSTEST}
\begin{figure}[!ht]

\centering
\includegraphics*[scale=0.42]{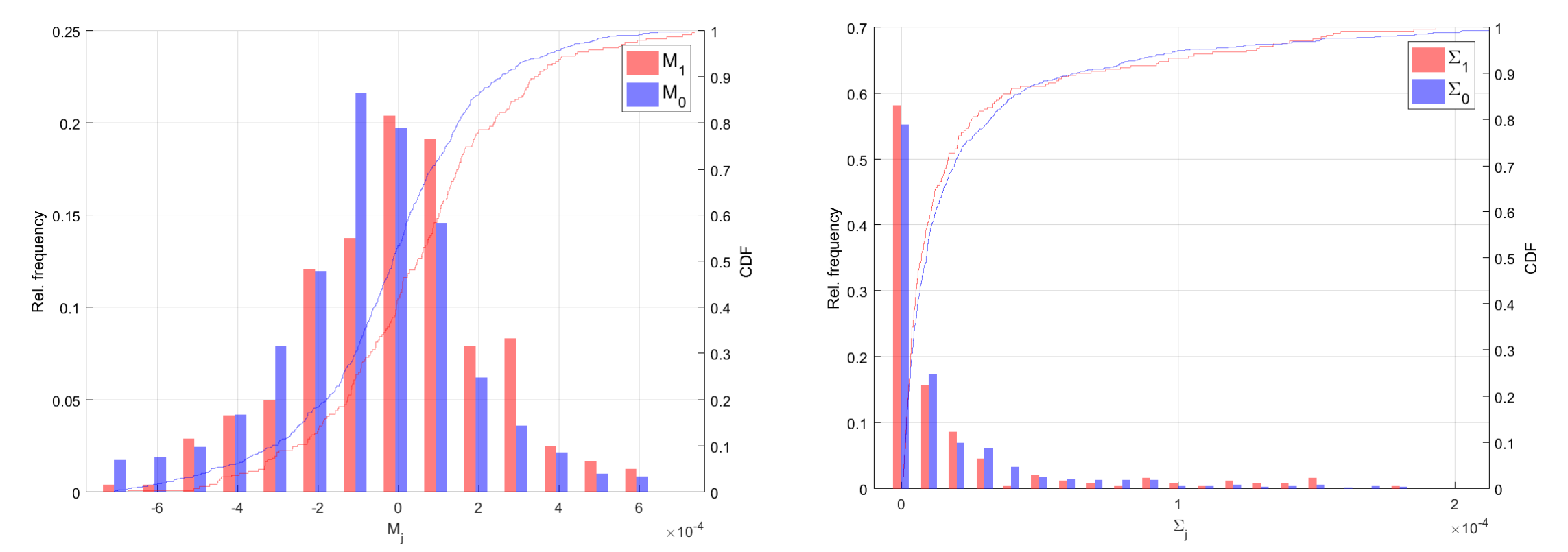}
\caption{Histograms and CDFs for $M_j$ and $\Sigma_j$. PC1 for $\Delta IV$, $\tau = 6$ months, 5 min windows for jump detection.}
\label{FIG:varianceDistr}

\end{figure}

First, as an illustrative example, we comment the left panel in Figure \ref{FIG:varianceDistr}, referring to distributions of $M_j$ for PC1 at six months maturity. Starting with the histograms, even if the two distributions seem to share the same degree of variability, $M_1$ looks slightly shifted towards positive values with respect to $M_0$ and exhibits right kurtosis. From the empirical CDF the dominance of $M_1$ is evident, so the higher probabilities of exceeding a given value and the slightly higher mean value.
This provides visual evidence that the distributions are indeed different and that higher means are more likely to be observed in the daily scores of the first hour in days with jumps in the underlying price. Since the first component corresponds to the OTM call options, this translates in observing (on average) higher changes in the levels of the smiles on its right side, where $m>1$ (larger change in IV for OTM calls, on average).

The qualitative encouragement provided by Figure \ref{FIG:varianceDistr} that the two samples are not drawn from the same distribution can be statistically verified by the Kolmogorov-Smirnov two-sample test. 
We first we consider the KS test with the most common hypotheses aimed at assessing whether a generic difference in the two distributions is statistically confirmed:

\begin{itemize}
\item $\mathbf{H_0}$:  $\forall u : F_{M_1}(u) =  F_{M_0}(u)$ (or simply $F_{M_1} = F_{M_0}$)
\item $\mathbf{H_1}$:  $\exists u : F_{M_1}(u) \neq  F_{M_0}(u)$ (or simply $F_{M_1} \neq F_{M_0}$),
\end{itemize}
\noindent where $F$ refers to the cumulative distribution. On the other, the most interesting scenarios are those where there's a dominance relationship between the CDFs: we test for the distribution $F_{M_1}$ to be uniformly \textit{greater} that the distribution $F_{M_0}$.

\begin{itemize}
\item $\mathbf{H^g_0}$:  $\exists u : F_{M_1}(u) \leq  F_{M_0}(u)$ (or simply $F_{M_1} \leq F_{M_0}$)
\item $\mathbf{H^g_1}$:  $F_{M_1}(u) >  F_{M_0}(u), \, \forall u$ (or simply $F_{M_1} > F_{M_0}$)
\end{itemize}
Analogously for the other tail, we test for $F_{M_1}$ being uniformly \textit{smaller} that the distribution of the scores variance in the morning of days without jumps $M_0$ ($F_{M_1}$ strictly dominates $F_{M_0})$:

\begin{itemize}
\item $\mathbf{H^s_0}$:  $\exists u : F_{M_1}(u) \geq  F_{M_0}(u)$ (or simply $F_{M_1} \geq F_{M_0}$)
\item $\mathbf{H^s_1}$:  $F_{M_1}(u) <  F_{M_0}(u), \, \forall u$ (or simply $F_{M_1} < F_{M_0}$)
\end{itemize}

\noindent The tests are conducted on the distributions $F_{M_j}$ and $F_{\Sigma_j}$, $j=0,1$ for all the maturities and all three principal components for $\Delta IV$ (differences in the IV smiles between consecutive times) with 5 min detection window (15 min window was also implemented for robustness checking). Table \ref{TAB:KS_NEW} we reports the p-values for all the three null hypotheses:

\begin{table}[!ht]
\centering
\scalebox{0.9}{

\begin{tabular}{cccccccccccc}
\toprule
&    &  \multicolumn{3}{c}{PC1}               & \multicolumn{3}{c}{PC2}               & \multicolumn{3}{c}{PC3}             \\
\cmidrule(l){3-5} \cmidrule(l){6-8}\cmidrule(l){9-11}
Maturity & Sign &  $H_0$ & $H^s_0$ & $H^g_0$ &  $H_0$ & $H^s_0$ & $H^g_0$ &  $H_0$ & $H^s_0$ & $H^g_0$\\
\hline 
\addlinespace[1ex] \multicolumn{1}{l}{\textbf{Test for $F_{M_j}$}}\\
3 months & pos  & 0.0816 	& 0.0408   & 0.9611   & 0.0000		& 0.0000   & 0.7926   & 0.0001	   & 0.0001   & 0.7531  \\
6 months & pos  & 0.0004	& 0.0002   & 1.0000   & 0.0000 		& 0.0000   & 0.4358   & 0.0000     & 0.0000   & 0.8581  \\
9 months & pos  & 0.0000	& 0.0000   & 0.8048   & 0.0000 		& 0.0000   & 0.5213   & 0.0000     & 0.0000   & 0.6199  \\
\addlinespace[1ex] \multicolumn{1}{l}{\textbf{Test for $F_{\Sigma_j}$}}\\
3 months & pos  & 0.7759 	& 0.7767   & 0.4181   & 0.0026		& 0.0013   & 0.6939   & 0.0740	   & 0.0370   & 0.8134  \\
6 months & pos  & 0.3232	& 0.8402   & 0.1623   & 0.0000 		& 0.0000   & 0.9967   & 0.0003     & 0.0002   & 0.8503  \\
9 months & pos  & 0.9044	& 0.7431   & 0.5254   & 0.0000 		& 0.0000   & 0.7555   & 0.0000     & 0.0000   & 0.9651  \\
\bottomrule
\end{tabular}}
\caption{Two sample KS test p-values for $F_{M_j}$ and $F_{\Sigma_j}$, $j = 0,1$. Five minutes jump detection window.}
\label{TAB:KS_NEW}
\end{table}

In interpreting the results in Table \ref{TAB:KS_NEW} we must take into consideration the sign of the loadings of each component, since it may reverse the interpretation of the results. In Table \ref{TAB:KS_NEW} the \lq\lq Sign\rq\rq column indeed reports the (general) sign of the loading for each PC, this can be visually double-checked from parallel coordinates plots like those in Figure \ref{fig:loadings}. In case of positive loading, accepting $H^s_1$ ($H^g_1$) means that a smaller (greater) CDF of $M_1$ with respect to $M_0$ translates in a higher (lower) mean change of the smile. For negative component loadings, the interpretation is reversed.

Overall the test on the mean scores of the mornings shows that the distributions of $M_j$ are statistically different and that $H^s_0$ is always rejected with high significance. Having evidence for $F_{M_1} < F_{M_0}$ and being the loadings for all the PCs positive we conclude that jumps in the underlying shift the distribution of the morning scores towards higher positive values, leading to larger quantiles and in general (and on average) to higher (positive) changes in IV and thus to higher IV throughout all the smile domain. 

Note that apparently less significant p-values as for the KS test on $F_{M_j}$ for PC1 and $\tau=3$ in \ref{TAB:KS_NEW} may arise from peculiar situations, e.g. where the CDF are very close and cross each other but with a very large portion of domain where one CDFs locally dominates the other. 

The very same analysis is proposed for the distributions $F_{\Sigma_0}$ and $F_{\Sigma_1}$. For the PC1, the right panel of Figure \ref{FIG:varianceDistr} depicts how the two distributions look alike and Table \ref{TAB:KS_NEW} reports analogous p-values for the hypotheses described above. Interestingly, we observe that the variance of the scores seems not to be affected by jumps in the underlying for the first component (corresponding to call options) while for the other components $H^s_1$ is generally accepted. Considering the positive sign of the components, Table \ref{TAB:KS_NEW} indicates that the variability of the IV smile changes is increasing for $m \leq 1$ (OTM puts and ATM call and puts) around jumps in contrast to the normal level, and this holds for any maturity.

Finally, all the conclusions here discussed for both $M_j$ and $\Sigma_j$ are consistent and unchanged while switching to 15 minutes detection window for jumps in the underlying price. The results are available upon request.

\subsection{Two-Sample Welch-U test}\label{SC:WELCHU}
In the previous section the distributions $F_{M_j}$ and $F_{\Sigma_j}$ were analysed. Here we rather focus on their first moment and test whether the means of $M_1$ and $\Sigma_1$ are respectively different from the respective means of $M_0$ and  $\Sigma_0$ (in presence of jumps versus no jumps). Even if the dominance in the CDF in the hypothesis $H^s_1$ implies a higher mean for $M_1$ the problem is non trivial, especially in those cases where the CDF cross each other. Moreover now we aim to address an additional analysis and conduct an explicit test on the mean values of the distributions.

A common tool for assessing the difference in means of two samples is the well-known t-test, however from Figure \ref{FIG:varianceDistr} we observe that the distributions of interest are clearly non-normal and that variances are unequal. This break the assumptions of the parametric t-test. \cite{zimmerman1993rank} address the issue of non-normality and heteroscedasticity in the samples for the parametric significance of the t-test, showing that the Welch-t test on the ranks simultaneously counteracts the effects of non-normality and non-equality of variances. Therefore, we compare the means of $M_j$ and $\Sigma_j$ with the Welch-t test performed on their respective ranks, which we refer to with Welch-U test.

To compute the mean values of the ranks, we first determine the ranks of the vector obtained merging $M_0$ and $M_1$ and create two distinct sets: one collecting the ranks of the elements coming from $M_0$ the other collecting the ranks of the elements originally members of $M_1$. We call the mean values of these subsets $\mu_{M_0}$ and $\mu_{M_1}$ respectively. Analogously $\mu_{\Sigma_j}$ denote the mean values of the ranks of $\Sigma_j$, $j = 0,1$.

We first consider the two-sided version of the test and then we conduct its one-tailed counterparts to understand which tail is (possibly) breaking $H_0$:

\begin{multicols}{3}

\begin{itemize}
\item $\mathbf{H_0}$: $\mu_{M_1} = \mu_{M_0}$ 
\item $\mathbf{H_1}$: $\mu_{M_1} \neq \mu_{M_0}$ 
\end{itemize}

\begin{itemize}
\item $\mathbf{H^s_0}$: $\mu_{M_1} \geq \mu_{M_0}$  
\item $\mathbf{H^s_1}$: $\mu_{M_1} < \mu_{M_0}$ 
\end{itemize}

\begin{itemize}
\item $\mathbf{H^g_0}$: $\mu_{M_1} \leq \mu_{M_0}$  
\item $\mathbf{H^g_1}$: $\mu_{M_1} > \mu_{M_0}$ 
\end{itemize}

\end{multicols}

In Table \ref{TAB:WELCH_NEW} we report the p-values for the three hypotheses above (which are analogous for $\mu_{\Sigma_j}$). Focusing first on the results for $\mu_{M_j}$, the p-values for the Welch-t test on ranks show that for all the maturities and for all the components $H_0$ is always rejected. Most interestingly, at the same time we always observe that $H^g_1$ is rejected (mostly with very high significance) and coherently $H^s_0$ accepted. Given the positive sign of the loadings, the interpretation of the Welch-U p-values is straightforward: the mean value of the ranks of morning scores for days with jumps in the underlying is (strictly) larger than in days without jumps.

When analysing the means of $\Sigma_1$ against $\Sigma_0$, for the first component, we do not observe any statistical difference in the means $\mu_{\Sigma_j}, j = 0,1$. For the other components there is evidence supporting $H^g_1$ for $\tau = 6, \, 9$, but for $\tau = 3$ no difference in the means is observed at all. The results show that for OTM call options (PC1) there are no differences in the mean variance of the scores when jumps occur in the morning, however for OTM puts and ATM options the mean variance of the scores is higher with large significance in morning with jumps in comparison to morning without jumps, but only for 6 and 9 months maturities. Results with 15 min detection window are coherent with the above results (available upon request). 

\begin{table}[!ht]
\centering
\scalebox{0.9}{

\begin{tabular}{cccccccccccc}
\toprule
&    &  \multicolumn{3}{c}{PC1}               & \multicolumn{3}{c}{PC2}               & \multicolumn{3}{c}{PC3}             \\
\cmidrule(l){3-5} \cmidrule(l){6-8}\cmidrule(l){9-11}
Maturity & Sign &  $H_0$ & $H^s_0$ & $H^g_0$ &  $H_0$ & $H^s_0$ & $H^g_0$ &  $H_0$ & $H^s_0$ & $H^g_0$\\
\hline 
\addlinespace[1ex] \multicolumn{1}{l}{\textbf{Test for $\mu_{M_j}$}}\\
3 months & pos  & 0.0856 	& 0.9572   & 0.0428   & 0.0009		& 0.9995   & 0.0005   & 0.0008	   & 0.9996   & 0.0004  \\
6 months & pos  & 0.0003	& 0.9999   & 0.0001   & 0.0008 		& 0.9996   & 0.0004   & 0.0002     & 0.9999   & 0.0001  \\
9 months & pos  & 0.0011	& 0.9995   & 0.0005   & 0.0007 		& 0.9997   & 0.0003   & 0.0008     & 0.9996   & 0.0004  \\
\addlinespace[1ex] \multicolumn{1}{l}{\textbf{Test for $\mu_{\Sigma_j}$}}\\
3 months & pos  & 0.7171 	& 0.3585   & 0.6415   & 0.0937		& 0.9532   & 0.0468   & 0.2687	   & 0.8657   & 0.1314  \\
6 months & pos  & 0.2188	& 0.1059   & 0.8941   & 0.0000 		& 1.0000   & 0.0000   & 0.0000     & 1.0000   & 0.0000  \\
9 months & pos  & 0.6855	& 0.3428   & 0.6572   & 0.0000 		& 1.0000   & 0.0000   & 0.0000     & 1.0000   & 0.0000  \\
\bottomrule
\end{tabular}}
\caption{Welch-U test p-values for $\mu_{M_j}$ and $\mu_{\Sigma_j}$, $j = 0,1$. Five minutes jump detection window.}
\label{TAB:WELCH_NEW}
\end{table}

Note that the Welch-U test represents an alternative to the t-test for non-normality and heteroschedasticity, in a way that Welch-U test can represent a proxy for the difference in means that one would usually test with the parametric t-test. This is, however, not the same thing. The difference is tiny, but any conclusion is therefore not referred to the mean values of the distributions, but more precisely to their ranks. Finally note that all the results here discussed are coherent with the dominance in CDFs discussed in the previous section, in particular for the variance of the scores note that when there were not differences in the CDFs $\Sigma_j$ we coherently have observed no differences in the mean ranks $\mu_{\Sigma_j}$.

\subsection{Discussion on the findings}\label{SUBSEC:DISCUSS}

We analyzed the behavior of the IV smile for three volatility slices from the scores of the first three principal components, each of which has a precise interpretation. We studied the differences in the distributions of the scores and the differences of their first moment assessing the direction of all the inequalities with the Kolmogorov-Smirnov and Welch-U tests respectively. The p-values in Table \ref{TAB:KS_NEW} and \ref{TAB:WELCH_NEW} have clear interpretation and the positive sign of the loadings for each of the three components reveals how the different hypotheses impact the IV smile. Results are briefly summarized in Table \ref{TAB:DISCUSS}.

\begin{table}[!ht]
\centering
\scalebox{0.9}{

\begin{tabular}{lcccc}
\toprule
&  \multicolumn{2}{c}{Mean($\Delta IV$)}  &  \multicolumn{2}{c}{Var($\Delta IV$)} \\
\cmidrule(l){2-3} \cmidrule(l){4-5}
& $\tau = 3$ & $\tau = 6,9$ & $\tau = 3$ & $\tau = 6,9$\\
\hline 
\addlinespace[1ex] \multicolumn{1}{l}{\textbf{Out-of-money puts}}\\
\, \, KS test 		& + & + & + & +\\
\, \, Welch-U test 	& + & + & = & +\\
\addlinespace[1ex] \multicolumn{1}{l}{\textbf{At-the-money puts and calls}}\\
\, \, KS test 		& + & + & + & +\\
\, \, Welch-U test 	& + & + & = & +\\
\addlinespace[1ex] \multicolumn{1}{l}{\textbf{Out-of-money calls}}\\
\, \, KS test 		& + & + & = & =\\
\, \, Welch-U test 	& + & + & = & =\\
\bottomrule
\end{tabular}}
\caption{Effect of jumps on $\Delta IV$. \lq\lq +\rq\rq \, indicates $F_{M_1} < F_{M_0}$ ($F_{\Sigma_1} < F_{\Sigma_0}$) or $\mu_{M_1} > \mu_{M_0}$ ($\mu_{\Sigma_1}> \mu_{\Sigma_0}$). \lq\lq =\rq\rq \, indicates no statistical evidence of any difference in $M_j$ ($\Sigma_j$) or $\mu_{M_j}$ ($\mu_{\Sigma_j}$).}
\label{TAB:DISCUSS}
\end{table}

It has been observed that the distributions of $M_j$ are different through all the maturities and that $M_1 $dominates $ M_0$. As commented earlier this results in a shift of the IV smile. Note that for $\tau=3$, the p-value of the Kolmogorov-Smirnov test for the PC1 is significant at 10\%, which indicates a less evident vertical shift of the surface. The distributions $F_{\Sigma_j}$, $j =1,2$ are statistically different for OTM puts and ATM options but not for OTM calls, however a difference arises in the means $\mu_{\Sigma_j}$ across different maturities. In particular, for the shortest maturity we observe that even though $F_{\Sigma_1} \neq F_{\Sigma_0}$ their respective means (on ranks) remain unchanged, this appears not to be the case for $\tau=6,9$, where the distribution of the scores show higher variance than normal but without differences in (rank) means.

\section{Conclusion}\label{SEC:CONCL}

The main purpose of this work was to investigate how the behavior of the implied volatility smile around jumps differs from its behavior when there are no jumps in the underlying price process, contributing to the literature with a first glimpse into the high-frequency analysis of the implied volatility dynamics with intra-day option data. \\
This was achieved by characterizing the surface with the first three principal components of the implied volatility smiles at three different maturities, showing each component to have a clear interpretation in terms of option moneyness. 
We focused on the changes in implied volatility in the first hour of the trading day, considering mean and variances of the morning scores deducted from PCA of mornings with jumps and not. The two score samples were separately studied in terms of similarity in their distributions and mean values, yielding further understanding of the differences and similarities in the behavior of the smile in the presence of jumps. 

On a general level, all the analyses we addressed, suggest a remarkable interconnection between jumps in the price of the underlying and implied volatility dynamics. According to our analysis, a wide and sudden change in the price of the underlying (jump) leads to a different implied volatility dynamics. This contradicts the independent assumption commonly used in stochastic volatility, jump-diffusion option pricing models, providing incentive to develop pricing models able to capture the observed linkage. 

We found that the distribution of the scores were statistically different for all the PCs and maturities we considered between the morning with jumps and not. We studied the direction of this asymmetry pointing out that this corresponds to abnormal vertical shifts of the IV surface as well as abnormal high variability through all the moneyness domain. \noindent Our findings show that impact of jumps on the implied volatility smile is strictly dependent on the moneyness and option type. In particular, for the out-of-money call options the implied volatility dynamics was observed not to be different in terms of scores variance whether we observe jumps in the underlying or not. On the other hand, we found at-the-money options and out-of-money puts to be highly reactive to jumps in the underlying.

Further research in this topic should strive more sophisticated methods for characterizing the analysis also in the direction of maturity and such as the functional principal component analysis could be employed in expanding this study.

\section*{Acknowledgements}

This project has received funding from the European Union’s Horizon 2020 research and innovation
programme under the Marie Skłodowska-Curie grant agreement No 675044.

\newpage

\bibliographystyle{apalike}	
\bibliography{Article_pca}

\begin{thebibliography}{}

\bibitem[Balland, 2002]{balland2002}
Balland, P. (2002).
\newblock Deterministic implied volatility models.
\newblock {\em Quantitative Finance}, 2(1):31--44.

\bibitem[Barndorff-Nielsen and Shephard, 2004]{barndorff2004power}
Barndorff-Nielsen, O.~E. and Shephard, N. (2004).
\newblock Power and bipower variation with stochastic volatility and jumps.
\newblock {\em Journal of financial econometrics}, 2(1):1--37.

\bibitem[Bates, 1996]{bates1996jumps}
Bates, D.~S. (1996).
\newblock Jumps and stochastic volatility: Exchange rate processes implicit in
  deutsche mark options.
\newblock {\em Review of financial studies}, 9(1):69--107.

\bibitem[Benko et~al., 2009]{benko2009}
Benko, M., H{\"a}rdle, W., and Kneip, A. (2009).
\newblock Common functional principal components.
\newblock {\em The Annals of Statistics}, 37(1):1--34.

\bibitem[Bernales and Guidolin, 2015]{bernales2015}
Bernales, A. and Guidolin, M. (2015).
\newblock Learning to smile: Can rational learning explain predictable dynamics
  in the implied volatility surface?
\newblock {\em Journal of Financial Markets}, 26:1--37.

\bibitem[Canina and Figlewski, 1993]{canina1993informational}
Canina, L. and Figlewski, S. (1993).
\newblock The informational content of implied volatility.
\newblock {\em Review of Financial studies}, 6(3):659--681.

\bibitem[Christensen and Prabhala, 1998]{christensen1998relation}
Christensen, B.~J. and Prabhala, N.~R. (1998).
\newblock The relation between implied and realized volatility.
\newblock {\em Journal of Financial Economics}, 50(2):125--150.

\bibitem[Cont and da~Fonseca, 2002]{cont2002}
Cont, R. and da~Fonseca, J. (2002).
\newblock Dynamics of implied volatility surfaces.
\newblock {\em Quantitative finance}, 2(1):45--60.

\bibitem[Duffie et~al., 2000]{duffie2000transform}
Duffie, D., Pan, J., and Singleton, K. (2000).
\newblock Transform analysis and asset pricing for affine jump-diffusions.
\newblock {\em Econometrica}, 68(6):1343--1376.

\bibitem[Dumas et~al., 1998]{dumas1998}
Dumas, B., Fleming, J., and Whaley, R.~E. (1998).
\newblock Implied volatility functions: Empirical tests.
\newblock {\em The Journal of Finance}, 53(6):2059--2106.

\bibitem[Fengler, 2006]{fengler2006}
Fengler, M.~R. (2006).
\newblock {\em Semiparametric modeling of implied volatility}.
\newblock Springer Science \& Business Media.

\bibitem[Fengler et~al., 2003]{fengler2003}
Fengler, M.~R., H{\"a}rdle, W.~K., and Villa, C. (2003).
\newblock The dynamics of implied volatilities: A common principal components
  approach.
\newblock {\em Review of Derivatives Research}, 6(3):179--202.

\bibitem[Gatheral, 2011]{gatheral2011volatility}
Gatheral, J. (2011).
\newblock {\em The volatility surface: a practitioner's guide}, volume 357.
\newblock John Wiley \& Sons.

\bibitem[Heynen, 1994]{heynen1994empirical}
Heynen, R. (1994).
\newblock An empirical investigation of observed smile patterns.
\newblock {\em Review of Futures Markets}, 13:317--317.

\bibitem[Heynen et~al., 1994]{heynen1994}
Heynen, R., Kemna, A., Vorst, T., et~al. (1994).
\newblock Analysis of the term structure of implied volatilities.
\newblock {\em Journal of Financial and Quantitative Analysis}, 29(1).

\bibitem[Homescu, 2011]{homescu2011}
Homescu, C. (2011).
\newblock Implied volatility surface: Construction methodologies and
  characteristics.
\newblock {\em Available at SSRN 1882567}.

\bibitem[Izenman, 2008]{izenman2008}
Izenman, A.~J. (2008).
\newblock {\em Modern multivariate statistical techniques}, volume~1.
\newblock Springer.

\bibitem[Lamoureux and Lastrapes, 1993]{lamoureux1993forecasting}
Lamoureux, C.~G. and Lastrapes, W.~D. (1993).
\newblock Forecasting stock-return variance: Toward an understanding of
  stochastic implied volatilities.
\newblock {\em Review of Financial Studies}, 6(2):293--326.

\bibitem[Lee and Mykland, 2008]{lee2008jumps}
Lee, S.~S. and Mykland, P.~A. (2008).
\newblock Jumps in financial markets: A new nonparametric test and jump
  dynamics.
\newblock {\em Review of Financial studies}, 21(6):2535--2563.

\bibitem[Lin et~al., 2008]{lin2008smiling}
Lin, B.-H., Chang, I.-J., and Paxson, D.~A. (2008).
\newblock Smiling less at liffe.
\newblock {\em Journal of Futures Markets}, 28(1):57--81.

\bibitem[Panigirtzoglou and Skiadopoulos, 2004]{panigirtzoglou2004}
Panigirtzoglou, N. and Skiadopoulos, G. (2004).
\newblock A new approach to modeling the dynamics of implied distributions:
  Theory and evidence from the s\&p 500 options.
\newblock {\em Journal of Banking \& Finance}, 28(7):1499--1520.

\bibitem[Rubinstein, 1994]{rubinstein1994implied}
Rubinstein, M. (1994).
\newblock Implied binomial trees.
\newblock {\em The Journal of Finance}, 49(3):771--818.

\bibitem[Skiadopoulos, 2001]{skiadopoulos2001}
Skiadopoulos, G. (2001).
\newblock Volatility smile consistent option models: a survey.
\newblock {\em International Journal of Theoretical and Applied Finance},
  4(03):403--437.

\bibitem[Skiadopoulos et~al., 2000]{skiadopoulos2000}
Skiadopoulos, G., Hodges, S., and Clewlow, L. (2000).
\newblock The dynamics of the s\&p 500 implied volatility surface.
\newblock {\em Review of Derivatives Research}, 3(3):263--282.

\bibitem[Wahba, 1990]{wahba1990spline}
Wahba, G. (1990).
\newblock {\em Spline models for observational data}, volume~59.
\newblock Siam.

\bibitem[Xu and Taylor, 1994]{xu1994magnitude}
Xu, G. and Taylor, S. (1994).
\newblock The magnitude of implied volatility smiles: Theory and empirical
  evidence for exchange rates.
\newblock {\em Review of Futures Markets}, 13:355--380.

\bibitem[Yang and Kanniainen, 2016]{kanniain2016jump}
Yang, H. and Kanniainen, J. (2016).
\newblock Jump and volatility dynamics for the s\&p 500: Evidence for
  infinite-activity jumps with non-affine volatility dynamics from stock and
  option markets.
\newblock {\em Review of Finance (to appear)}.

\bibitem[Zhu and Avellaneda, 1997]{zhu1997}
Zhu, Y. and Avellaneda, M. (1997).
\newblock An e-arch model for the term structure of implied volatility of fx
  options.
\newblock {\em Applied Mathematical Finance}, 4:81--100.

\bibitem[Zimmerman and Zumbo, 1993]{zimmerman1993rank}
Zimmerman, D.~W. and Zumbo, B.~D. (1993).
\newblock Rank transformations and the power of the student t test and welch
  t'test for non-normal populations with unequal variances.
\newblock {\em Canadian Journal of Experimental Psychology/Revue canadienne de
  psychologie exp{\'e}rimentale}, 47(3):523.

\end{thebibliography}

\end{document}